\documentclass[12pt]{article}
\usepackage{geometry}
\usepackage{graphicx}
\usepackage{amsmath}
\usepackage{amsfonts}
\usepackage{amssymb}
\begin{document}
	\title{Radiation from an accelerating charge in a family of Rindler frames}
	\author{Luther Rinehart}
	\maketitle
	\begin{abstract}
		The generalization of the Larmor radiation formula in gravitational fields and with accelerating observers was obtained by Hirayama and others. We verify a special case of their result by explicit computation using a family of displaced Rindler frames. We discuss the role of observer-dependence of energy and simultaneity. We also include a discussion of conservation laws in spacetimes equipped with a Killing vector and a time function.
	\end{abstract}
	\section{Background}
	The relationship between acceleration (Larmor) radiation and the equivalence principle of relativity is an old paradox in classical electromagnetism. The main questions are of two types:
	\begin{enumerate}
		\item A charged particle with nonzero proper acceleration may be interpreted by a comoving observer as being at rest in a gravitational field. Does such an observer observe the emission of radiation?
		\item A charged particle undergoing inertial motion may be interpreted by an accelerating observer as being in free fall in a gravitational field. Does such an observer observe the emission of radiation?
	\end{enumerate}
	Both questions remain controversial, with many explanations and thought experiments being proposed. See \cite{Gron} for a review. Nevertheless, many regard the authoritative resolution to have been given by Rohrlich \cite{Rohrlich}, who answered no to question 1 and yes to question 2. If we set aside issues of conceptual interpretation, both questions may be framed as well-posed problems of classical field theory, and answered by explicit calculation of the electromagnetic field in the appropriate coordinate system. The result is the solution given by Rohrlich. \par
	In this classical field theory context, the resolution of the apparent paradox is that the notion of radiation is observer-dependent. As will be reflected in the following calculations, this observer-dependence arises from two sources. The first is the observer's notion of energy, which depends on their state of motion in a manner well-understood even at the level of special relativity. For non-inertial motions and coordinate systems, and if the energy is distributed over a region of space, the observer's instantaneous velocity is no longer sufficient to define their notion of energy. In general spacetimes, what is needed for this is a timelike Killing vector field, generating the observer's notion of time-translation symmetry. \par
	The second source of observer-dependence of radiation is the observer's notion of simultaneity. Radiation, as a \emph{rate} of energy, also depends on how the observer does their time-slicing, or what events they regard as simultaneous. For inertial motions, the relativities of energy and simultaneity cancel each other under Lorentz transformations, and that is why the rate of radiation is Lorentz-invariant. But as shown by the calculations of Rohrlich, the cancellation does not persist for more general motion. These two inputs, stationarity and simultaneity, are both needed to define radiation. They are given by the data of a Killing field and a time function. We see that the concept of an observer cannot simply be quantified by the observer's worldline, but requires as input a frame of reference extending throughout a spacetime region.\par
	Kretzschmar and Fugmann \cite{KF1,KF2}, as well as Hirayama \cite{Hirayama1,Hirayama2}, have given the generalization of the Larmor formula to the case where observer and charge have arbitrary velocities and accelerations. The classic questions about the equivalence principle and radiation in accelerated frames may be treated as special cases of this. Suppose the particle has velocity $v^\mu$ and proper acceleration $a^\mu$, and suppose the worldlines of the observer's reference frame have velocity $u^\mu$ and proper acceleration $g^\mu$. Let $g=\sqrt{g_\mu g^\mu}$. Let
	\begin{equation}
	h^\mu_{\ \nu} = \delta^\mu_{\ \nu} + v^\mu v_\nu
	\end{equation}
	be the projection orthogonal to $v^\mu$. Define the Hirayama acceleration vector as
	\begin{equation}\label{Hirayama}
	\alpha^\mu = h^\mu_{\ \nu}(a^\nu-g^\nu-gu^\nu)
	\end{equation}
	All these quantities are to be evaluated at the retarded time, that is, at the point of emission. The general result is that the radiated power is proportional to $\alpha^\mu\alpha_\mu$.\par
	The purpose of this paper is to reproduce this result, via explicit calculation, for the case where $v^\mu=u^\mu$ and $a^\mu$ and $g^\mu$ are collinear. At the time of performing this calculation, the author was unaware of the work of Kretzschmar, Fugmann, and Hirayama. \par
	Rindler coordinates describe how Minkowski spacetime looks to a uniformly accelerated observer. The main technical tool of this paper is to use a family of Rindler coordinate systems, parameterized by their $z$-displacement, and by their proper acceleration. This way, the possibilities for displacement between observer and charge, and for their two independent accelerations, are at least partly parameterized. 
	\section{The field and stress-energy of a point charge in arbitrary motion}
	This section consists of standard textbook material which can be found, for example, in \cite{Rbook} chapter 5. We will use signature $-,+,+,+$, and employ dot product notation for the four-dimensional dot product. That is, $A\cdot B = A_\mu B^\mu$. Consider a particle of charge $q$ in Minkowski spacetime, with four-velocity $v^\mu$ and proper four-acceleration $a^\mu$. Let $R^\mu = x^\mu - x^\mu_0$ be the four-dimensional displacement vector, where $x^\mu_0$ is the location of the particle. The Lienard-Wiechert potential of the point charge in arbitrary motion is
	\begin{equation}
	A_\nu = \frac{qv_\nu}{4\pi v\cdot R}
	\end{equation}
	where everything is evaluated at the retarded time, or to put it another way, on the future lightcone, characterized by $R\cdot R=0$. The field tensor is
	\begin{equation}
	F_{\mu\nu} = \frac{q}{4\pi(v\cdot R)^2}\left[ R_\mu(a_\nu-fv_\nu)- R_\nu(a_\mu-fv_\mu)\right]
	\end{equation}
	where 
	\begin{equation}
	f = \frac{1+a\cdot R}{v\cdot R}
	\end{equation}
	The stress-energy tensor is
	\begin{equation}
	T_{\mu\nu}=\frac{q^2}{(4\pi)^2(v\cdot R)^4} \Big[R_\mu R_\nu(a^2-f^2) +R_\mu(a_\nu-fv_\nu) +R_\nu(a_\mu-fv_\mu) + \frac{1}{2}g_{\mu\nu}\Big]
	\end{equation}
	The acceleration of the charge need not be uniform. The field on the future lightcone of an instant on the particle's worldline depends only on $v^\mu$ and $a^\mu$ at that instant. We may choose coordinates for the inertial frame so that the charge passes through the origin and is instantaneously at rest, with its acceleration in the $z$-direction. Also in the inertial frame, let 
	\begin{equation}
	x_\bot = \sqrt{x^2 +y^2}
	\end{equation}
	\begin{equation}
	r=\sqrt{x_\bot^2 +z^2}
	\end{equation}
	With the inertial frame coordinates so chosen, we have
	\begin{align}
	&R^\mu = r\hat{t} +z\hat{z} +x_\bot\hat{x}_\bot\\
	&a^\mu = a\hat{z}\label{a}\\
	&v^\mu = \hat{t}\label{v}\\
	&v\cdot R = -r\\
	&a\cdot R = az
	\end{align}
	The parameter $a$ may be positive or negative. The idea is to fix the origin as the point of emission we are interested in. We will look at the field on the future light cone of the origin. The behavior of the charge at other points of its trajectory does not matter in this context, because we are using a massless field. The emission point is fixed at the origin, and the observer's coordinate system will be moved around.
	\section{A family of displaced Rindler frames}
\begin{figure}[h]
		\centering
		\includegraphics[width=3.25in]{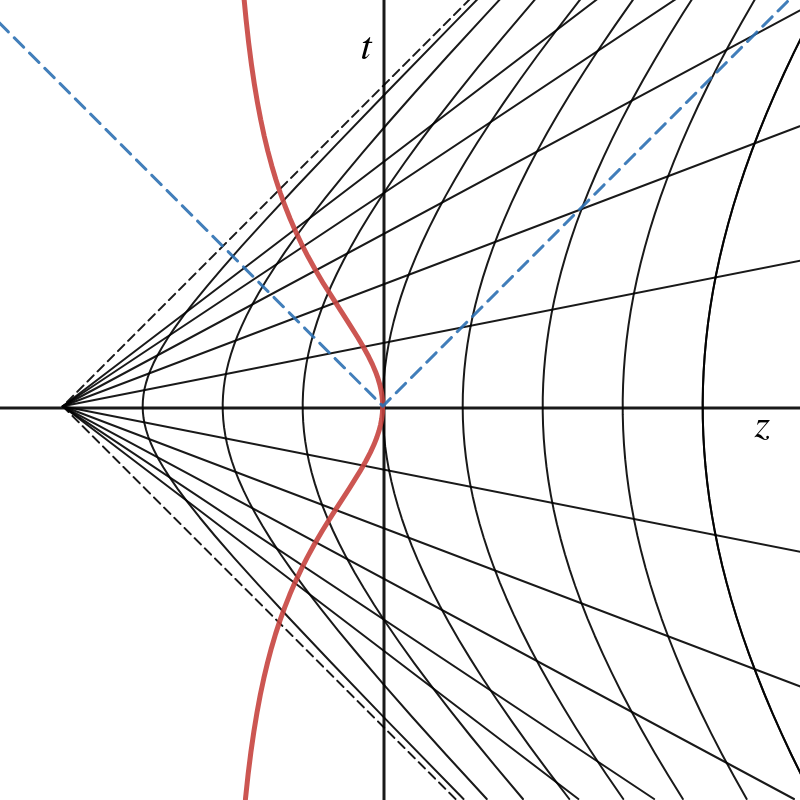}
		\caption{The shifted Rindler coordinate system, with an example source trajectory (red), and the future lightcone of the origin (blue)}
		\label{fig:shifted}
	\end{figure}

	Introduce a family of Rindler frames with coordinates $\chi$ and $\tau$, and parameters $\chi_0$ and $g$, taken to be positive. The coordinates are related to the inertial coordinates by 
	\begin{equation}
	t=\frac{\chi}{g}\sinh(g\tau)\qquad z+\frac{\chi_0}{g} = \frac{\chi}{g}\cosh(g\tau)
	\end{equation}
	and cover the wedge of spacetime where $z+\frac{\chi_0}{g} >|t|$. This setup is depicted in figure \ref{fig:shifted}. In these coordinates, the metric is
	\begin{equation}
	g_{\mu\nu} = -\chi^2d\tau^2 +\frac{d\chi^2}{g^2}+dx^2+dy^2
	\end{equation}
	representing a uniform gravitational field. Rindler time $\tau$ is a time coordinate for which the metric is invariant. Among $\tau$-stationary observers, those at $\chi=1$ are privileged in that their proper time coincides with the Rindler time $\tau$. The worldlines of $\tau$-stationary observers are in uniform acceleration, although with differing values of proper acceleration. Privileged observers at $\chi=1$ have proper acceleration equal to $g$. The origin of the inertial coordinate system, where the charge is located, in the Rindler coordinates is at $\chi=\chi_0$. Thus $\chi_0$ represents the vertical displacement of the charge in the Rindler spacetime, and also represents a redshift factor between the observer and the charge. \par
	The $\tau$ symmetry of Rindler spacetime is associated with a Killing field 
	\begin{equation}
	\xi^\mu = g\left( z+\frac{\chi_0}{g} \right)\hat{t} + g t\hat{z}
	\end{equation}
	which gives us the Rindler observers' notion of conserved energy. \par
\begin{figure}[h]
		\centering
		\includegraphics[height=3.25in]{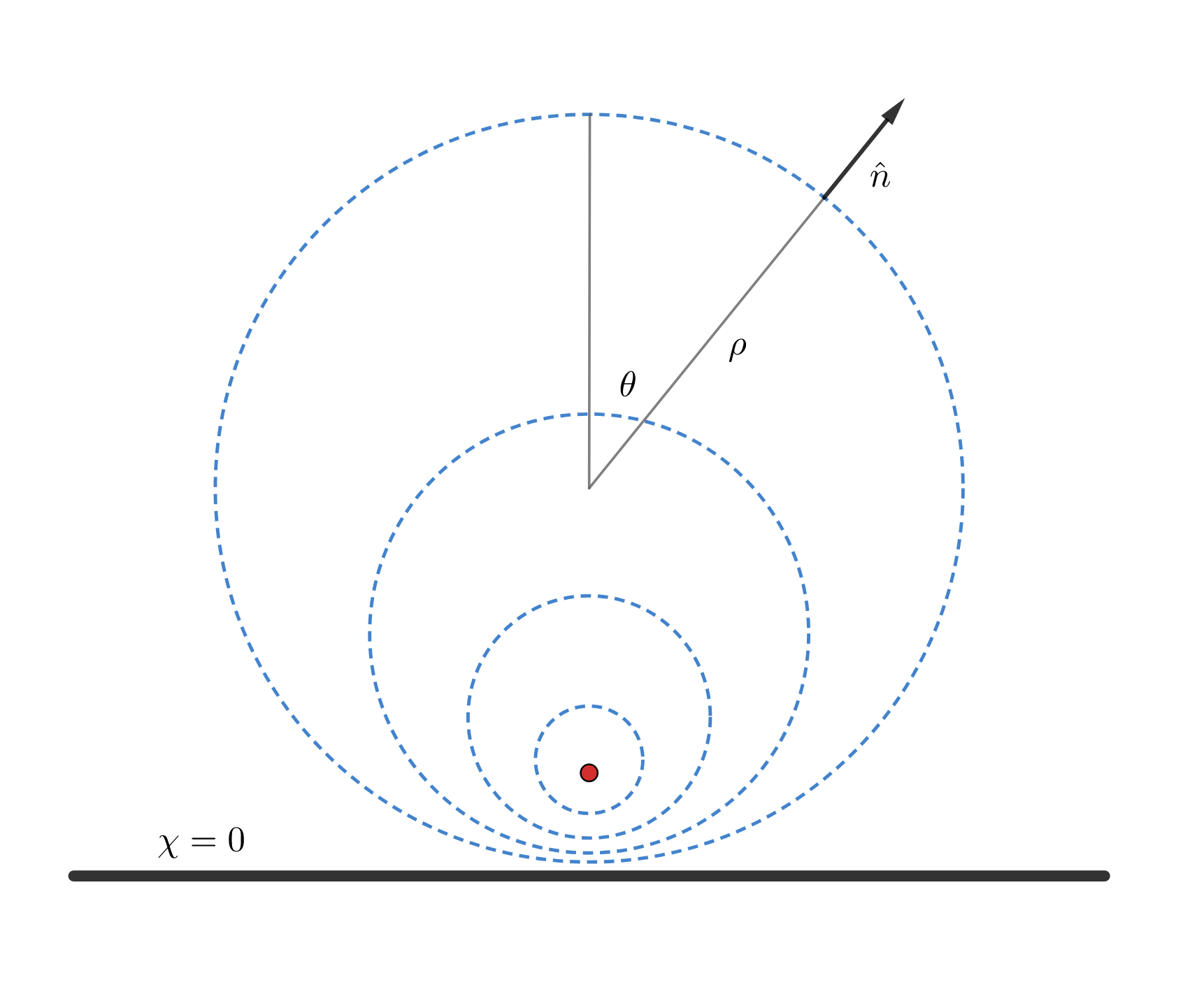}
		\caption{The future lightcone of the origin at successive times $\tau$ as seen in Rindler space, shown in the $x$-$\chi$ plane}
		\label{fig:lightspheres}
	\end{figure}

	The 3-dimensional geometry of a constant $\tau$ hypersurface is Euclidean. The future lightcone of the origin intersected with such a surface turns out to be a sphere centered at 
	\begin{equation}
	\chi=\chi_0\cosh(g\tau)
	\end{equation}
	with radius
	\begin{equation}
	\rho = \frac{\chi_0}{g}\sinh{g\tau}
	\end{equation}
	On this sphere,
	\begin{equation}
	t=r=\frac{\chi\rho}{\chi_0}
	\end{equation}
	The geometry of the lightcone sphere as seen in Rindler space is shown in figure \ref{fig:lightspheres}.  We may introduce a polar angle $\theta$ on the lightcone sphere, and express the other coordinates in terms of $\rho$ and $\theta$:
	\begin{align}
	&x_\bot =\rho\sin\theta\\
	&z=\frac{g\rho}{\chi_0}\left( \rho + \sqrt{\rho^2+\frac{\chi_0^2}{g^2}}\cos\theta \right)\\
	&r=\frac{g\rho}{\chi_0}\left(\sqrt{\rho^2+\frac{\chi_0^2}{g^2}}+\rho\cos\theta \right)
	\end{align}
	The unit normal vector to the lightcone sphere is 
	\begin{equation}
	\begin{split}
	\hat{n} &= \cos\theta\,\hat{\chi} + \sin\theta\,\hat{x}_\bot\\
	&= \frac{g\rho\cos\theta}{\chi_0}\,\hat{t} +\frac{g\cos\theta}{\chi_0}\sqrt{\rho^2+\frac{\chi_0^2}{g^2}}\,\hat{z} +\sin\theta\,\hat{x}_\bot
	\end{split}
	\end{equation}
	\section{Evaluating the Rindler energy flux on the light-sphere}
	We may now calculate the flux of Rindler energy through the lightcone sphere as
	\begin{equation}
	S = -T_{\mu\nu}\xi^\mu\hat{n}^\nu
	\end{equation}
	The needed ingredients are the following dot products:
	\begin{align}
	&R\cdot\xi = -r\chi_0\\
	&R\cdot \hat{n}=\rho\\
	&\hat{n}\cdot \xi =0\\
	&(a-fv)\cdot\hat{n} = \frac{\rho\cos\theta}{r\chi_0}(a\chi_0-g)\\
	&(a-fv)\cdot \xi = -\frac{1}{r}(g z +\chi_0 +az\chi_0 -ag x^2_\bot)
	\end{align}
	Putting these together in the stress tensor gives
	\begin{equation}
	\begin{split}
	S= -\frac{q^2}{(4\pi)^2r^4}\Big[ &-\frac{\rho\chi_0}{r}(a^2x_\bot^2-2az-1)-\rho\cos\theta(a\chi_0-g)\\
	&-\frac{\rho}{r}(g z +\chi_0 + az\chi_0 -ag x_\bot^2)\Big]\\
	=-\frac{q^2\rho}{(4\pi)^2r^5}&(a\chi_0-g)(-ax_\bot^2 +z-r\cos\theta)\\
	=\frac{q^2\rho}{(4\pi)^2r^5}&(a\chi_0-g)\left(ax_\bot^2 -\frac{g\rho^2\sin^2\theta}{\chi_0}\right)\\
	=&\frac{q^2\rho^3\sin^2\theta}{(4\pi)^2\chi_0r^5}(a\chi_0-g)^2
	\end{split}
	\end{equation}
	At this point several qualitative features are apparent. 
	\begin{itemize}
		\item In the $g\rightarrow 0$ limit, the radiation has the same features as Larmor radiation. 
		\item An inertial charge, with $a=0$, does radiate for accelerated observers. 
		\item A Rindler-stationary charge is one with $a\chi_0=g$, and does not radiate.
	\end{itemize}
	Recall that $a$ may be positive or negative, accounting for whether the observer and charge accelerate in the same or opposite direction. However $\chi_0$ must be positive, since otherwise the future lightcone of the origin does not intersect the observer's Rindler wedge. 
	\section{Digression on the interpretation of the conservation equation}\label{conservation1}
	In arbitrary spacetimes, a conservation law is expressed locally by the equation
	\begin{equation}
	\nabla_\mu J^\mu =0
	\end{equation}
	Suppose we have a static metric of the form
	\begin{equation}
	g_{\mu\nu}=-u(x)dt^2 +g^{(3)}_{\mu\nu}(x)
	\end{equation}
	where $g^{(3)}$ is a metric on 3-dimensional space independent of $t$. The Rindler spacetime is of this form, but also others like Schwarzschild. The vector field $J^\mu$ decomposes as $(J^t,\overrightarrow{J})$. Then the conservation equation takes the form
	\begin{equation}
	\frac{\partial}{\partial t}J^t + \frac{1}{\sqrt{u}}\nabla\cdot (\sqrt{u}\overrightarrow{J})=0
	\end{equation}
	where the second term involves the 3-dimensional divergence with respect to $g^{(3)}$. This can be rewritten
	\begin{equation}
	\frac{\partial}{\partial t}(\sqrt{u}J^t) + \nabla\cdot (\sqrt{u}\overrightarrow{J})=0
	\end{equation}
	The interpretation is that the conserved quantity has density $\sqrt{u}J^t$ and flux $\sqrt{u}\overrightarrow{J}$. The volume factor $\sqrt{u}$ must be included, even though it does not appear in $g^{(3)}$. This result is sufficient for the purpose of the present Rindler spacetime calculation. It is generalized in section \ref{conservation2} to the case where the Killing vector need not be hypersurface-orthogonal.
	\section{Results}
	In Rindler spacetime, the volume factor is $\chi$, making the flux 
	\begin{equation}
	\chi S = \frac{q^2\rho^2\sin^2\theta}{(4\pi)^2r^4}(a\chi_0-g)^2
	\end{equation}
	This may now be integrated over the light-sphere to give the power
	\begin{equation}
	\begin{split}
	P&= \int \chi S\ \rho^2d\Omega\\
	&= \frac{q^2(a\chi_0-g)^2}{(4\pi)^2} \int_0^\pi \frac{\rho^4}{r^4}\sin^2\theta\, 2\pi\sin\theta\ d\theta\\
	&= \frac{q^2(a\chi_0-g)^2}{8\pi}\int_0^\pi \left( \sqrt{1+\left(\frac{g\rho}{\chi_0}\right)^2}+\frac{g\rho}{\chi_0}\cos\theta \right)^{\!-4}\sin^3\theta\,d\theta
	\end{split}
	\end{equation}
	The $\theta$ integral is equal to $4/3$ for all values of $\rho$, so the final result is independent of $\rho$:
	\begin{equation}
	\begin{split}
	P&= \frac{2}{3}\ \frac{q^2(a\chi_0-g)^2}{4\pi}\\
	&= \frac{2}{3}\ \frac{q^2\chi_0^2}{4\pi}\left( a-\frac{g}{\chi_0} \right)^2
	\end{split}
	\end{equation}
	Note that, when integrated over the sphere, the radiation rate is independent of the distance from the charge. This is in agreement with the results of Rohrlich \cite{Rohr2}. This is true even retaining all the terms in the energy flux, and not employing any large-distance, radiation limit.\par
	We may interpret this result in the language of Hirayama as in \eqref{Hirayama}, and verify that it is in agreement with the general formula found in \cite{Hirayama1,Hirayama2}. $a^\mu$ and $v^\mu$ are as given in equations \eqref{a} and \eqref{v}. In the Rindler frame, the $\tau$-stationary worldlines have velocity $u^\mu = \frac{1}{\chi}\xi^\mu$, and proper acceleration
	\begin{equation}
	g^\mu = u^\nu\nabla_\nu u^\mu = \frac{g^2}{\chi^2}\left( t\hat{t} +(z+\chi_0/g)\hat{z}\right)
	\end{equation}
	These must be evaluated at the origin, giving $u^\mu=\hat{t}$ and $g^\mu = \frac{g}{\chi_0}\hat{z}$. Since $u^\mu = v^\mu$, the Hirayama acceleration vector simplifies to 
	\begin{equation}
	\alpha^\mu = a^\mu - g^\mu = \left(a-\frac{g}{\chi_0}\right)\hat{z}
	\end{equation}
	and the radiation is indeed proportional to $\alpha^\mu\alpha_\mu$. The factor of $\chi_0^2$ may be interpreted as a redshift factor between the source, located at $\chi=\chi_0$, and the observer, located at $\chi=1$. The power is doubly redshifted to account for both the redshift of the energy, and also the time dilation.
	
	\section{Conservation laws in a spacetime with a Killing vector and a time function}\label{conservation2}
	This section generalizes the result of section \ref{conservation1} in a coordinate-independent manner, dropping the assumption that the Killing vector be hypersurface-orthogonal. If a spacetime is equipped with a stationary frame of reference in the form of a Killing vector and a time function, this structure allows the equation $\nabla_\mu J^\mu=0$ to be interpreted by stationary observers as a local conservation law for some conserved quantity, in a manner compatible with the observers' notions of time-translation and simultaneity. Consider a spacetime equipped with a timelike Killing vector field $\xi^\mu$ and a time function $t$ that is compatible in the sense that its Lie derivative is equal to 1:
	\begin{equation}
	\mathsterling_\xi(t)=1
	\end{equation}
	In particular, this means
	\begin{equation}
	\mathsterling_\xi(\nabla_\mu t)=0
	\end{equation}
	so the foliation of constant-$t$ hypersurfaces is preserved under the transformation generated by $\xi^\mu$. This data represents a notion of time-translation and simultaneity. It allows us to interpret spacetime as spatial manifold $\Sigma$ with a time evolution. Let $h_{\mu\nu}$ be the restriction of the metric to the constant-$t$ surface $\Sigma$. $h_{\mu\nu}$ is preserved under time evolution, and as a Riemannian metric, endows $\Sigma$ with its own geometry and covariant derivatives. We seek to interpret the conservation equation in a manner compatible with this picture.\par
	An arbitrary vector field $J^\mu$ uniquely decomposes as 
	\begin{equation}
	J^\mu = J^0\xi^\mu + j^\mu
	\end{equation}
	where
	\begin{equation}
	J^0=J^\mu\nabla_\mu t
	\end{equation}
	\begin{equation}
	j^\mu\nabla_\mu t=0
	\end{equation}
	meaning that $j^\mu$ is tangent to the surfaces of constant $t$.
	We may then calculate
	\begin{equation}
	\nabla_\mu J^\mu = \mathsterling_\xi J^0 + \nabla_\mu j^\mu
	\end{equation}
	However, the second term is not the same as the divergence of $j^\mu$ with respect to the metric $h_{\mu\nu}$ on $\Sigma$. The derivative operator $D$ on $\Sigma$ associated with $h_{\mu\nu}$ is the orthogonal projection onto $\Sigma$ of the operator $\nabla$. Hence the divergence we are interested in is
	\begin{equation}
	D_\mu j^\mu = \nabla_\mu j^\mu + n_\mu n^\nu\nabla_\nu j^\mu
	\end{equation}
	where $n^\mu$ is the unit normal vector to $\Sigma$. Using the fact that $j^\mu n_\mu=0$, this equals
	\begin{equation}
	= \nabla_\mu j^\mu - j^\mu n^\nu\nabla_\nu n_\mu
	\end{equation}
	Now introduce a function $f$ defined by
	\begin{equation}
	n_\mu =f\nabla_\mu t
	\end{equation}
	Now we have
	\begin{equation}
	\begin{split}
	D_\mu j^\mu &= \nabla_\mu j^\mu - j^\mu n^\nu\nabla_\nu (f\nabla_\mu t)\\
	&= \nabla_\mu j^\mu - j^\mu n^\nu f\nabla_\nu \nabla_\mu t\\
	\end{split}
	\end{equation}
	Since $\nabla$ is torsion-free, this equals
	\begin{equation}
	\begin{split}
	&= \nabla_\mu j^\mu - j^\mu n^\nu f\nabla_\mu \nabla_\nu t\\
	&= \nabla_\mu j^\mu - j^\mu n^\nu f\nabla_\mu\left(\frac{n_\nu}{f}\right) 
	\end{split}
	\end{equation}
	Finally, since $n^\nu n_\nu =-1$,
	\begin{equation}
	\begin{split}
	&= \nabla_\mu j^\mu + j^\mu f\nabla_\nu \left(1/f\right)\\
	&= f\nabla_\mu \left(\frac{j^\mu}{f}\right)
	\end{split}
	\end{equation}
	This result holds for arbitrary vector fields tangent to $\Sigma$. Hence all together,
	\begin{equation}
	\nabla_\mu J^\mu = \mathsterling_\xi J^0 + \frac{1}{f}D_\mu (fj^\mu)
	\end{equation}
	Note that 
	\begin{equation}
	\frac{1}{f}= n^\mu\nabla_\mu t = \sqrt{\nabla_\mu t \nabla^\mu t}
	\end{equation}
	and consequently, $\mathsterling_\xi (f)=0$. The conservation equation can be rewritten
	\begin{equation}
	0 = \mathsterling_\xi(fJ^0) + D_\mu (fj^\mu)
	\end{equation}
	By using Gauss's law in $\Sigma$, we have the interpretation that there is a conserved quantity of density $fJ^0$ and flux $fj^\mu$.
\section{Summary and conclusions}
Applying the equivalence principle to the Larmor radiation of accelerating charged particles leads naturally to questions of how classical acceleration radiation looks to accelerated observers, and in gravitational fields. The work of Rohrlich \cite{Rohrlich}, in answering such questions, revealed that the concept of radiation is observer-dependent, and opened the way for a generalization of the Larmor formula to account for the view of non-inertial observers. Such generalization was given by Kretzschmar and Fugmann \cite{KF1,KF2}, and Hirayama \cite{Hirayama1,Hirayama2}. To define a total rate of radiation, an observer must be equipped with a reference frame extending throughout a region of spacetime, in the form of a timelike Killing vector field, and a time-slicing. The result is that the observed radiated power is proportional to the square of the \emph{difference} between the proper acceleration of the source, and the proper acceleration of the frame. In particular, an inertial charge is seen to radiate by an accelerated observer, and an accelerating charge is not seen to radiate by a co-accelerated observer. In this paper, we have attempted to show, by explicit calculation, how this result comes about, and how it depends on both the observer's notion of energy, given by the Killing vector, and the observer's notion of simultaneity, given by the time-slicing.\par
The general formula given by Hirayama was derived in the case where the observer accelerates uniformly in a flat background. However, its terms could easily be interpreted in any spacetime equipped with a Killing vector and a time function, such as the Schwarzschild spacetime. For example, a charged particle freely orbiting a black hole would presumably radiate to stationary observers, despite having vanishing proper acceleration. The author is not aware of any work to rigorously justify a generalization of the Hirayama formula to other spacetimes. Therefore such generalization, if possible, is an opportunity for future work.
\section{Acknowledgments}
Collaborative discussion was provided by Stephen Fulling of Texas A\&M University, Gerard Kennedy of the University of Southampton, and Timothy Bates, student at Texas A\&M.	
		
\end{document}